\documentclass[aps,prd,twocolumn,floatfix,nofootinbib,showpacs,superscriptaddress,tightenlines]{revtex4}

\usepackage{amsmath}
\usepackage{dcolumn}
\usepackage{lipsum}
\usepackage{amssymb}
\usepackage{url}
\usepackage{epsfig}
\usepackage{graphicx}
\usepackage{amsmath}
\usepackage{bm}
\usepackage{comment} 
\usepackage{setspace}
\usepackage{lscape}
\usepackage{amsthm}
\usepackage{bbold}
\usepackage{dcolumn}
\usepackage{epsfig}
\usepackage{graphics}
\usepackage{graphicx}
\usepackage{longtable}
\usepackage{color}
\usepackage{xcolor}
\usepackage{bm}
\usepackage{xspace}
\usepackage{cancel}
\usepackage{float}
\usepackage{multirow}
\usepackage[mathscr,scaled=1.15]{urwchancal}
\DeclareFontFamily{OT1}{pzc}{}
\DeclareFontShape{OT1}{pzc}{m}{it}%
{<-> s * [1.15] pzcmi7t}{}
\DeclareMathAlphabet{\mathpzc}{OT1}{pzc}{m}{it}

\definecolor{darkgreen}{rgb}{0,0.5,0}
\definecolor{amber}{rgb}{1.0, 0.75, 0.0}
\definecolor{purple}{rgb}{0.5,0,0.5}
\definecolor{nblue}{rgb}{0.0,0.0,0.50}
\definecolor{scarlet}{rgb}{1.0,0.2,0}
\definecolor{darkmagenta}{rgb}{0.55, 0.0, 0.55}
\definecolor{darkolivegreen}{rgb}{0.33, 0.42, 0.18}
\definecolor{darkcandyapplered}{rgb}{0.64, 0.0, 0.0}
\definecolor{warmblack}{rgb}{0.0, 0.26, 0.26}
\definecolor{oxfordblue}{rgb}{0.0, 0.13, 0.28}
\definecolor{cyan(process)}{rgb}{0.0, 0.55, 0.55}
\definecolor{almond}{rgb}{0.94, 0.87, 0.8}
\definecolor{antiquewhite}{rgb}{0.98, 0.92, 0.84}
\definecolor{eggshell}{rgb}{0.94, 0.92, 0.84}
\definecolor{floralwhite}{rgb}{1.0, 0.98, 0.94}
\definecolor{linen}{rgb}{0.98, 0.94, 0.9}

\usepackage[colorlinks=true, pdfstartview=FitV, linkcolor=purple, citecolor= purple, urlcolor=blue]{hyperref}

\newcommand{\be}{\begin{equation}}
\newcommand{\tu}{\textcolor{red}{u}}

\newcommand{\fu}{\textcolor{blue}{\bar{f_2}}}
\newcommand{\fd}{\textcolor{blue}{f_1}}
\newcommand{\fdu}{\textcolor{blue}{f_2}}

\newcommand{\Me}{\textcolor{blue}{V}}
\newcommand{\Meps}{\textcolor{blue}{{PS}}}
\newcommand{\Dps}{\textcolor{blue}{{DPS}}}
\newcommand{\Mv}{\textcolor{blue}{V}}
\newcommand{\Mav}{\textcolor{blue}{AV}}
\newcommand{\Ms}{\textcolor{blue}{S}}
\newcommand{\Ds}{\textcolor{blue}{{DS}}}
\newcommand{\Dv}{\textcolor{blue}{{DV}}}
\newcommand{\Dav}{\textcolor{blue}{{DAV}}}

\newcommand{\Jpsi}{\textcolor{blue}{J/\Psi}}

\newcommand{\td}{\textcolor{darkcandyapplered}{d}}
\newcommand{\tb}{\textcolor{blue}{b}}
\newcommand{\tc}{\textcolor{darkmagenta}{c}}
\newcommand{\ts}{\textcolor{darkgreen}{s}}
\newcommand{\ee}{\end{equation}}
\newcommand{\bea}{\begin{eqnarray}}
\newcommand{\eea}{\end{eqnarray}}
\newcommand{\beas}{\begin{eqnarray*}}
\newcommand{\eeas}{\end{eqnarray*}}
\newcommand{\nn}{\nonumber}

\newcommand{\MeV}{\text{MeV}} 
\newcommand{\GeV}{\text{GeV}} 
\newcommand{\rmh}{\hat{\alpha}_{\mathrm {IR}}}

\begin{document}
\title{First Radial Excitations of Mesons and Diquarks in a Contact Interaction}

\author{G. Paredes-Torres}
\email[]{gustavo.paredes@umich.mx}
\thanks{}
\affiliation{Instituto de F\'isica y Matem\'aticas, Universidad
Michoacana de San Nicol\'as de Hidalgo, Edificio C-3, Ciudad
Universitaria, Morelia, Michoac\'an 58040, M\'exico}

\author{L.X. Guti\'errez-Guerrero}
\email[]{lxgutierrez@conacyt.mx}
\thanks{}
\affiliation{Instituto de F\'isica y Matem\'aticas, Universidad
Michoacana de San Nicol\'as de Hidalgo, Edificio C-3, Ciudad
Universitaria, Morelia, Michoac\'an 58040, M\'exico}
\affiliation{CONACyT-Mesoamerican Centre for Theoretical Physics,
Universidad Aut\'onoma de Chiapas, Carretera Zapata Km. 4, Real
del Bosque (Ter\'an), Tuxtla Guti\'errez 29040, Chiapas, M\'exico}

\author{A. Bashir}
\email[]{adnan.bashir@umich.mx}
\thanks{}
\affiliation{Instituto de F\'isica y Matem\'aticas, Universidad
Michoacana de San Nicol\'as de Hidalgo, Edificio C-3, Ciudad
Universitaria, Morelia, Michoac\'an 58040, M\'exico}
\affiliation{Dpto. Ciencias Integradas, Centro de Estudios Avanzados en Fis., Mat. y Comp.,
Fac. Ciencias Experimentales, Universidad de Huelva, Huelva 21071, Spain}

\author{\'Angel S. Miramontes}
\thanks{}
\email[]{angel.miramontes@umich.mx}
\affiliation{Instituto de F\'isica y Matem\'aticas, Universidad
Michoacana de San Nicol\'as de Hidalgo, Edificio C-3, Ciudad
Universitaria, Morelia, Michoac\'an 58040, M\'exico}
%

\begin{abstract}
We present a calculation for the masses of the first radially excited states of forty mesons and diquarks made up of $\tu,\td,\ts,\tc$ and $\tb$ quarks, including states that contain one or both heavy quarks. To this end, we employ a combined analysis of the Bethe-Salpeter and Schwinger-Dyson equations within a self-consistent and symmetry preserving vector-vector contact interaction. 
The same set of parameters describe ground and excited states of mesons and their diquark partners. The wave-function of the first radial excitation contains a zero whose location is correlated with an additional parameter $d_F$ which is a function of dressed quark masses.   
Our results satisfy the equal spacing rules given by the Gell-Mann Okubo mass relations. Wherever possible, we make
comparisons of our findings with known experimental observations as well as theoretical predictions of
several other models and approaches including lattice quantum chromodynamics, finding a very good agreement.
We report predictions for a multitude of radial excitations not yet observed in experiments.
\end{abstract}


\maketitle

\section{Introduction}

A detailed exploration of the ground and excited states of two-body bound systems unravels the dynamics of the underlying quantum field theory. As a standout example, precise measurements of 
the energy spectrum of the Hydrogen atom, carried out by Lamb in 1947, led to the birth of renormalized quantum electrodynamics (QED). In a similar manner,
we expect meson spectroscopy to be essential to our understanding of dynamical chiral symmetry breaking (DCSB) and confinement, the emergent infrared phenomena which characterize quantum chromodynamics (QCD). 
However, a comprehensive insight is achieved not merely by reproducing the ground states satisfactorily. It also requires accurately describing their excitations.
Radially excited states, while retaining the same spin and parity as the ground states, naturally have higher masses. 
For example, first and second radial excitations of a pion, discovered experimentally almost 40 years ago~\cite{Bellini:1982ec}, have masses of $1300$ MeV and $1800$ MeV, respectively.
Since then, many theoretical efforts have been made to study these excitations of mesons which include approaches like the Bethe-Salpeter equation (BSE)~\cite{Holl:2004fr,Chang:2019eob,Chen:2012qr,Roberts:2011cf,Mojica:2017tvh,Li:2016dzv,Rojas:2014aka,Xu:2022kng}, lattice QCD~\cite{Gao:2021hvs,Mastropas:2014fsa,Dudek:2010wm,McNeile:2006qy}, QCD sum rules~\cite{Gelhausen:2014jea,Jiang:2015paa,Kataev:2004ve}, linear sigma model~\cite{Parganlija:2016yxq}, holographic QCD~\cite{Ballon-Bayona:2014oma,Krein:2016gua,Afonin:2020msa,Ahmady:2022dfv} and non-relativistic potential models~\cite{Alhakami:2019ait,Asghar:2019qjl,Gutierrez-Guerrero:2021fuj}.

    On the experimental front, several radial excitations of mesons have been reported over the years. Initial observations include the ones of the pion~\cite{Bellini:1982ec}, as mentioned before, the $\rho$ meson~\cite{Diekmann:1986wq} via pion-nucleus collisions at Serpukhov accelerator and CERN, identification and study of the radial excitation of the $K$ mesons by LHCb, ACCMOR collaboration, and SLAC,~\cite{LHCb:2017swu,ACCMOR:1981yww,Brandenburg:1976pg}, and the confirmation of the radial excitations of $D^0$, $D^{*0}$ and $D^{*+}$ mesons by BaBar experiment~\cite{BaBar:2010zpy}. 
    This experiment also studied whether the controversial $D^{*}_{\ts}(2710)$ was the first
radial excitation of $D^{*}_{\ts}(2112)$~\cite{BaBar:2009rro}.   
BESIII detector at the BEPCII $e^+e^-$
 collider provided precise mass determination of $\eta_{\tc}(1S)$\footnote{We employ the convention where $n=0$ corresponds to the ground state. We refer to the text for the detailed notation.} meson~\cite{BESIII:2013nja}. 
LHCb Collaboration \cite{LHCb:2015aaf} observed a state consistent with $B^{*}_{\tc}(1S)$ with a mass of $6841.2$ MeV.
The two excited $B_{\tc}$ mesons, $B^+_{\tc}(1S)$ and $B^{*+}_{\tc}(1S)$, were observed  for the first time
by the CMS experiment at $\sqrt{s}=13\textmd{ TeV}$~\cite{CMS:2019uhm}.
CLEO III reported the observation of $\eta_{\tb}(1S)$ \cite{Dobbs:2012zn} and carried out the first determination of the hyperfine mass splitting in the bottomonium sector: $\Upsilon$($1S$) - $\eta_{\tb}$($1S$). Precise mass measurements for $\Jpsi$ and $\Upsilon$ radial excitations have been performed at the VEPP-4 Collider~\cite{Artamonov:1983vz,OLYA:2000toq}. In the realm of scalar and axial vector mesons, we have limited experimental results to date, with observations of $D_{\ts 0}^*$ \cite{BaBar:2009rro}, $\chi_{\tb 0}$ \cite{BaBar:2014och}, $a_1$ \cite{DELPHI:1998bhv}, and $\chi_{\tb 1}$ \cite{CLEO-II:1991wqd} meson radial excitations. Note that a critical component of the 
 ongoing scientific program of the 12 GeV upgrade at the Thomas Jefferson National Accelerator Facility is the study of the spectrum and structure of excited hadrons~\cite{Arrington:2021alx}. This rather brief and incomplete survey of the experimental efforts should be enough to convey how
 active this field of study has been in the past and continues to be.

 The study of mesons through BSEs also serves as a portal to the the exploration of baryons, specially while employing a quark-diquark picture to explore their properties.  
 Diquarks are colour non-singlet quark-quark correlations. These diquark correlations are studied through the BSEs which are identical in form to the corresponding mesons with different charge and color factors. 
  Owing to their colour charge, like individual quarks and gluons, diquarks cannot propagate to the detectors. We assume this to be the case although the colored diquarks have an associated mass scale. Interest in diquarks has grown in recent years, thanks to their role and importance in calculating the properties of baryons, see for example~\cite{Barabanov:2020jvn}. In quark-diquark picture, two quarks in an attractive colour
anti-triplet configuration can couple with a third quark to form a colour singlet
baryon. Naturally, this approach alleviates the computational strenuousness since it only invokes a two body interaction instead of three~\cite{Maris:2004bp}.
Modern related
studies reaffirm that a full three-body equation yields nucleon masses
which do not differ from the quark-diquark picture by more than
$5\%$ \,\cite{Eichmann:2009qa}.
This quark-diquark picture  has been very successful in describing  masses \cite{Roberts:2011cf,Roberts:2011wy, Lu:2016bbk,Gutierrez-Guerrero:2021rsx,Yin:2021uom} and form factors \cite{Raya:2021pyr,Wilson:2011aa,Barabanov:2020jvn,Cui:2020rmu,Segovia:2016zyc} of baryons.
Scalar and vector-axial diquarks dominate in the nucleon with positive parity, while for baryons with negative parity the contributions of diquarks with both parities are indispensable \cite{Gutierrez-Guerrero:2021rsx,Yin:2021uom}.
In addition, diquarks are important in the study of tetraquarks and pentaquarks since these particles can be described as bound states containing diquarks and an additional quark in the latter case~\cite{Jaffe:2003sg,Faustov:2020qfm,Jin:2020jfc,Liu:2019zuc}. The
diquarks corresponding to their ground state mesons play a role for the relevant ground state baryons. However, it is conceivable that if we consider excited baryons, we would require to invoke excitations of diquarks. It establishes the relevance of considering the first radial excitation of the diquarks in this article.

 We investigate first radial excitations of mesons and diquarks through a coupled analysis of the BSEs for the two-body bound states and the Schwinger-Dyson equation (SDE) which triggers DCSB for a single quark in strongly coupled QCD. We employ
 a simple, efficient, symmetry preserving and self-consistent vector-vector contact interaction model (CI). It mimics infrared QCD to produce the mass spectrum of all ground state mesons and baryons by implementing essential features of infrared QCD such as confinement, DCSB and the low energy implications of axial vector Ward-Takahashi identity. Several works use CI to study mesons, see Table~\ref{CIstudies}.


 \begin{table*}[htbp]~\label{CIstudies}
 \renewcommand{\arraystretch}{1.6}
  \caption{The list of mesons that have been studied through a
symmetry-preserving SDE/BSE treatment of a vector-vector CI. \\}
        \begin{tabular}{@{\extracolsep{0.8 cm}}ll|ll}
             \hline\hline
            Pesudoscalar Mesons & Ref. & Scalar Mesons & Ref. \\ \rule{0ex}{3.5ex}
           $\pi(\tu\bar{\td})$ \cite{GutierrezGuerrero:2010md,Roberts:2011wy,Roberts:2011cf},
$K(\tu\bar{\ts})$,
$h_s(\ts\bar{\ts})$ &\cite{Chen:2016spr,Lu:2017cln,Gutierrez-Guerrero:2019uwa,Yin:2019bxe}& $\sigma(\tu\bar{\td})   $          \cite{Roberts:2011cf},
$K_0^*(\tu\bar{\ts})$, 
$f_0(\ts\bar{\ts})$&\cite{Chen:2016spr,Lu:2017cln}\\
\rule{0ex}{3.5ex}
$D^{0}(\tc\bar{\tu})$,
$D^{+}_{\ts}(\tc\bar{\ts})$,
$B^{+} (\tu\bar{\tb})$,
$B_s^0(\ts\bar{\tb})$
& \cite{Gutierrez-Guerrero:2019uwa,Yin:2019bxe} & $D_0^*(\tc\bar{\tu})$,
$D_{s0}^*(\tc\bar{\ts})$, 
$B_{0}^*(\tu\bar{\tb})$,
$B_{s0}(\ts\bar{\tb})$
 & \cite{Gutierrez-Guerrero:2021rsx,Yin:2021uom}  \\
\rule{0ex}{3.5ex}
$B_{\tc}^{+}(\tc\bar{\tb})$,$\eta_c(\tc\bar{\tc})$ \cite{Bedolla:2015mpa,Raya:2017ggu},
$\eta_{\tb}(\tb\bar{\tb})$ \cite{Raya:2017ggu} 
           & \cite{Gutierrez-Guerrero:2019uwa,Yin:2019bxe} & $B_{c0}(\tc\bar{\tb})$,$\chi_{c0}(\tc\bar{\tc})$          
             \cite{Bedolla:2015mpa,Raya:2017ggu},
$\chi_{b0}(\tb\bar{\tb})$ \cite{Raya:2017ggu}  & \cite{Gutierrez-Guerrero:2021rsx,Yin:2021uom} 
           \\
             \hline\hline
  Vector Mesons & Ref.  &  Axial-vector Mesons & Ref.           
             \\
 \rule{0ex}{3.5ex}
 $\rho(\tu\bar{\td})$ \cite{Roberts:2011wy,Roberts:2011cf},
$K_1$ $(\tu\bar{\ts})$,
$\phi$ $(\ts\bar{\ts})$ &\cite{Chen:2016spr,Lu:2017cln,Gutierrez-Guerrero:2019uwa,Yin:2019bxe}& $a_1(\tu\bar{\td})$ \cite{Roberts:2011cf},
$K_1(\tu\bar{\ts})$,
$f_1(\ts\bar{\ts})$&\cite{Chen:2016spr,Lu:2017cln}
\\
\rule{0ex}{3.5ex}
$D^{*0}(\tc\bar{\tu})$,
$D_{\ts}^{*}(\tc\bar{\ts})$, 
$B^{+*}(\tu\bar{\tb})$,
$B_{\ts}^{0*}(\ts\bar{\tb})$,
 &\cite{Gutierrez-Guerrero:2019uwa,Yin:2019bxe}&$D_1(\tc\bar{\tu})$,
$D_{s1}(\tc\bar{\ts})$,
$B_{1}(\tu\bar{\tb})$,
$B_{s1}(\ts\bar{\tb})$& \cite{Gutierrez-Guerrero:2021rsx,Yin:2021uom} 
\\
\rule{0ex}{3.5ex}
$B_{\tc}^{*}(\tc\bar{\tb})$,
$\Jpsi$ $(\tc\bar{\tc})$ \cite{Bedolla:2015mpa,Raya:2017ggu},
$\Upsilon(\tb\bar{\tb})$ \cite{Raya:2017ggu} &
\cite{Gutierrez-Guerrero:2019uwa,Yin:2019bxe}& 
$B_{cb}(\tc\bar{\tb})$,
$\chi_{c1}(\tc\bar{\tc})$ \cite{Bedolla:2015mpa,Raya:2017ggu},
$\chi_{b1}(\tb\bar{\tb})$ \cite{Raya:2017ggu}&  \cite{Gutierrez-Guerrero:2021rsx,Yin:2021uom} 
 \\
           \hline\hline 
\end{tabular}
    \label{tabIntro}
\end{table*}
This manuscript is organized as follows: in Sec.~\ref{CI-1}, we
recall general features of the CI, the underlying assumptions and a detailed discussion on how the parameters of the model are fitted. After providing a self-contained discussion on the BSE in~\ref{BSE-s},
our detailed analysis of the masses of the first radial excitation of mesons and diquarks can be found in subsections
\ref{Meson-radial} and \ref{diquark-radial}.  In Sec.~\ref{Conclusions}, we present our conclusions.
\section{ Contact interaction: features }
\label{CI-1}

For practical description of hadrons, the gap equation for the quarks requires modelling the gluon propagator and the quark-gluon vertex. 
In this section, we shall recall the main truncation and characteristics which define the CI~\cite{GutierrezGuerrero:2010md,Roberts:2010rn,Roberts:2011wy,Roberts:2011cf}~:
\begin{itemize}
\item  The gluon propagator is defined to be independent of any running momentum scale:
\begin{eqnarray}
\label{eqn:contact_interaction}
g^{2}D_{\mu \nu}(k)&=&4\pi\rmh\delta_{\mu \nu} 
\equiv
4\pi\frac{\alpha_{\mathrm {IR}}}{m_{g}^{2}}\delta_{\mu\nu} \,,
\end{eqnarray}
%
\noindent where $m_g=500\,\MeV$ is a gluon mass scale generated dynamically within infrared QCD~\cite{Boucaud:2011ug,Aguilar:2017dco,Binosi:2017rwj,Gao:2017uox} and $\alpha_{\mathrm{IR}}$ can be interpreted as the interaction strength in the infrared~\cite{Binosi:2016nme,Deur:2016tte,Rodriguez-Quintero:2018wma}.
\item We take the leading-order quark-gluon vertex:
\begin{equation}
\Gamma_{\nu}(p,q) =\gamma_{\nu} \,.
\end{equation}
\item With this kernel, the dressed-quark propagator for a quark of flavor $f$ becomes\footnote{We employ a Euclidean metric with:  $\{\gamma_\mu,\gamma_\nu\} = 2\delta_{\mu\nu}$; $\gamma_\mu^\dagger = \gamma_\mu$; $\gamma_5= \gamma_4\gamma_1\gamma_2\gamma_3$; and $a \cdot b = \sum_{i=1}^4 a_i b_i$.  A timelike four-vector, $Q$, has $Q^2<0$.  Furthemore, we consider isospin-symmetric limit.} 
\begin{eqnarray}
 \nn && S_f^{-1}(p) = \\
&&  i \gamma \cdot p + m_f +  \frac{16\pi}{3}\rmh \int\!\frac{d^4 q}{(2\pi)^4} \,
\gamma_{\mu} \, S_f(q) \, \gamma_{\mu}\,,\label{gap-1}
\end{eqnarray}
 where $m_f$ is the current-quark mass. The integral possesses quadratic and logarithmic divergences. We choose to regularize them in a Poincar\'e covariant manner. The solution of this equation is~: 
\begin{equation}
\label{genS}
S_f^{-1}(p) = i \gamma\cdot p + M_f \,,
\end{equation}
 where $M_f$ in general is the mass function running with a momentum scale. However, within the CI, it 
 is a constant dressed mass.
\item $M_f$ is determined by
\begin{equation}
M_f = m_f + M_f\frac{4\rmh}{3\pi}\,\,{\cal C}^{\rm iu}(M_f^2)\,,
\label{gapactual}
\end{equation}
where 
\bea 
\hspace{0.75 cm}
{\cal C}^{\rm iu}(\sigma)/\sigma = \overline{\cal C}^{\rm iu}(\sigma) = \Gamma(-1,\sigma \tau_{\rm uv}^2) - \Gamma(-1,\sigma \tau_{\rm ir}^2),
\eea
 $\Gamma(\alpha,y)$ is the incomplete gamma-function and $\tau_{\rm ir, uv}$ are respectively, infrared and ultraviolet regulators. A nonzero value for  $\tau_{\mathrm{IR}}\equiv 1/\Lambda_{\mathrm{IR}}$ implements 
confinement~\cite{Ebert:1996vx}. Since the CI is a nonrenormalizable theory, 
$\tau_{\mathrm{UV}}\equiv 1/\Lambda_{\mathrm{UV}}$ becomes part of the definition of the model and therefore sets the scale for
all dimensioned quantities. 
 \end{itemize}
 In this work we report results  using the parameter values detailed in Tables~\ref{parameters},~\ref{table-M}, which correspond to what were dubbed as 
 {\em heavy parameters} in~\cite{Gutierrez-Guerrero:2019uwa}. In this choice, the effective coupling constant and the ultraviolet regulator vary as a function of the quark mass. In the context of the CI\footnote{ For other models and the external conditions of finite density, temperature and/or magnetic field, similar ideas were earlier implemented in \cite{Casalbuoni:2003cs,Farias:2006cs,Farias:2014eca,Farias:2016let}.}, this behavior was first suggested in~\cite{Bedolla:2015mpa} and later adopted in several subsequent works~\cite{Bedolla:2016yxq,Raya:2017ggu,Serna:2017nlr,Gutierrez-Guerrero:2019uwa,Yin:2019bxe,Chen:2019otg,Yin:2021uom}. The heavier quarks 
 tend to make more compact mesons and the closer proximity of these quarks leads to lower values of the effective strong interaction coupling. This behavior is incorporated through the parameters choice in the third column of  Table~\ref{parameters}.
 
 \begin{table}[H]
 \caption{\label{parameters} Ultraviolet regulator (in GeV) as well as dimensionless coupling constant for different combinations of quarks in a hadron.  $\alpha_{\mathrm {IR}}=\alpha_{\mathrm{IRL}}/Z_H$ with $\alpha_{\mathrm {IRL}}=1.14$, extracted from the best-fit to data, as explained in~\cite{Raya:2017ggu}. Fixed parameters are the gluon mass $m_g =500$ MeV  reported in~\cite{Gao:2017uox} and $\Lambda_{\rm IR} = 0.24$ GeV.}
\begin{center}
\label{parameters1}
\begin{tabular}{@{\extracolsep{0.5 cm}}lccc}
\hline \hline
 quarks & $Z_H$ & $\Lambda_{\mathrm {UV}}\;[\GeV] $ & $\alpha_{\mathrm {IR}}$  \\
$\tu,\td,\ts$ & 1 & 0.905& 1.14\\
\rule{0ex}{2.5ex}
$\tc,\td,\ts$ & 3.034 & 1.322&0.376 \\
\rule{0ex}{2.5ex}
$\tc$     & 13.122 & 2.305&0.087 \\
\rule{0ex}{2.5ex}
$\tb,\tu,\ts$ & 16.473 & 2.522 & 0.069 \\
\rule{0ex}{2.5ex}
$\tb,\tc$   & 59.056 & 4.131 &0.019\\
\rule{0ex}{2.5ex}
$\tb$     & 165.848 & 6.559 & 0.007\\
\hline \hline
\end{tabular}
\end{center}
\end{table}
 %
%
%
Table~\ref{table-M} presents current quark masses in GeV used herein and the dressed masses of $\tu$, $\ts$, $\tc$ and $\tb$ quarks, computed from the gap equation, Eq.~(\ref{gapactual}).
\begin{table}[H]
\caption{\label{table-M}
Current ($m_{u,\cdots}$) and dressed masses 
($M_{u,\cdots}$) for quarks (GeV), required as input for the BSE and the FE.}
\vspace{0.3cm}
\begin{tabular}{@{\extracolsep{0.5 cm}}cccc}
\hline 
\hline
 $m_{\tu}=0.007$ &$m_{\ts}=0.17$ & $m_{\tc}=1.08$ & $m_{\tb}=3.92$   \\
 \rule{0ex}{3.5ex}
 $\hspace{-.2cm} M_{\tu}=0.367$ & $ \hspace{0.1cm} M_{\ts}=0.53$\; &  $ \hspace{0.0 cm} M_{\tc}=1.52$ & $ \hspace{0.0cm} M_{\tb}=4.68$  \\
 \hline
 \hline
\end{tabular}
\end{table}
The simplicity of the CI enables straightforward calculation of hadronic observables, including their masses, decay constants, charge radii and form factors. The study of heavy, heavy-light and light meson masses also leads to the evaluation of the mass scale associated with diquark correlations through using the same set of parameters. 
With this in mind, in the next section, we describe and solve the BSE for mesons and diquarks.
\section{Bethe Salpeter Equation}
\label{BSE-s}

\begin{figure}[htbp]
\vspace{-5.5cm}
       \centerline{
       \includegraphics[scale=0.5,angle=0]{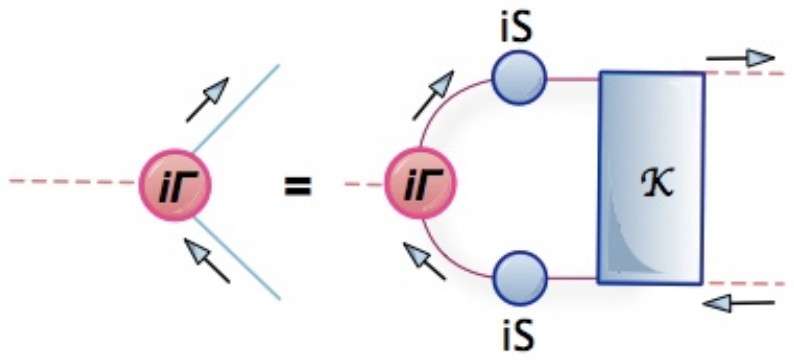}}
       \vspace{-5.5cm}
       \caption{\label{Fig0}This diagram represents the BSE. Blue (solid) circles represent dressed propagators $S$, red (solid) circle is the meson BSA $\Gamma$ and the blue (solid) rectangle is the dressed-quark-antiquark scattering kernel $K$.}
\end{figure}

The bound-state problem for hadrons characterized by two valence-fermions is studied using the
homogeneous BSE in Fig.~\ref{Fig0}.  
We explore how this equation describes mesons and diquarks in both their ground and excited states. The homogeneous BSE reads~\cite{Salpeter:1951sz},
%
\begin{equation}
[\Gamma(k;P)]_{tu} = \int \! \frac{d^4q}{(2\pi)^4} [\chi(q;P)]_{sr} K_{tu}^{rs}(q,k;P)\,,
\label{genbse}
\end{equation}
where $\Gamma$ is the bound-state's Bethe-Salpeter amplitude (BSA),
 $\chi(q;P) = S(q+P)\Gamma S(q)$ is the BS wave-function; $r,s,t,u$ represent colour, flavor and spinor indices and $K$ is the relevant fermion-fermion scattering kernel.  This equation possesses physical solutions at $P^2$-values for which bound-states exist.
We employ the notation introduced in~\cite{Chen:2012qr}, [$\fd,\fdu$] for scalar and pseudoscalar diquarks, and $(\{\fd,\fd\}),(\{\fd,\fdu\})$ for axial-vector and vector diquarks. We describe the details of the radially excited meson mass calculation in the following subsection.
\subsection{Radial excitations of mesons}
\label{Meson-radial}
To obtain the mass and amplitude associated
with the first radial excitation of a meson comprised of a quark with
flavor $\fd$ and an antiquark with flavor $\fu$,
 we employ the same methods as detailed in Refs.~\cite{Gutierrez-Guerrero:2019uwa,Gutierrez-Guerrero:2021rsx}. However, we naturally include an extra term associated with the fact that the first radial excitation possesses a single
zero just like the radial wave function for bound states in quantum mechanics. Within any sophisticated QCD based treatment of mesons, 
all Chebyshev moments of BSA in the first radial state possess a single zero,
whereas those of ground BSA
exhibit none \cite{Holl:2004fr}. 
In the phenomenological application of CI, we follow the works~\cite{Roberts:2011cf,Roberts:2011wy} and insert a zero by hand into the kernels in the BSE, multiplying it by $(1-q^2d_F)$
which forces a zero into the kernel at $q^2 = 1/d_F$ , where $d_F$ is an additional parameter. The presence of this zero
reduces the coupling in the BSE and hence increases the bound-state’s mass. The presence of this term modifies the functions ${\cal C}^{\rm iu}$ in \cite{Gutierrez-Guerrero:2021rsx}. It must now be replaced by ${\cal F}^{\rm iu} = {\cal C}^{\rm iu} - d_{\cal F} {\cal D}^{\rm iu}$ where
\begin{eqnarray}
\nonumber
{\cal D}^{\rm iu}(\omega) = \int_0^\infty ds\,s\,\frac{s}{s+\omega }
\to  \int_{\tau_{\rm uv}^2}^{\tau_{\rm iu}^2} d\tau\, \frac{2}{\tau^3} \,
\exp\left[-\tau \omega\right], \rule{1em}{0ex}
\end{eqnarray}
${\cal F}^{\rm iu}_1(z) = - z (d/dz){\cal F}^{\rm iu}(z)$ and $\overline{\cal F}_1(z) = {\cal F}_1(z)/z$.  The general decomposition of the bound-state's  BSA for radial excitations is the same as the ground state. In CI,
\bea
\label{BSA-mesones}
\Gamma_H=A_HE_H +B_H F_H\;,
\eea
where $H={\Meps, \Ms, \Mv, \Mav}$ denotes pseudoscalar ($\Meps$), scalar ($\Ms$), vector (\Mv),  and axial vector ($\Mav$) mesons, respectively.
The explicit form of the BSA  for
different types of mesons is displayed in Table~\ref{ff-BSE-1}.
 \begin{table}[htbp]
 \caption{\label{ff-BSE-1} We list BSAs for mesons. The total momentum of the bound state is $P$ and $M_R = M_{\fd} M_{\fu}/[M_{\fd} + M_{\fu}]$.}
\begin{center}
\begin{tabular}{@{\extracolsep{0.5 cm}}lcc}
\hline \hline
BSA & $A_H$ & $B_H$ \\
\rule{0ex}{2.0ex}
  $\Gamma_{\Meps}$ &$i\gamma_5$&$\frac{1}{2M_R}\gamma_5 \gamma\cdot P$ \\ 
  \rule{0ex}{2.0ex}
$\Gamma_{\Ms}$ &$I_D$ &-- \\ 
\rule{0ex}{2.0ex}
$\Gamma_{\Mv,\mu}$& $\gamma_{\mu}^{T}$&$\frac{1}{2M_R}\sigma_{\mu\nu}P_\nu$ \\ 
\rule{0ex}{2.5ex}
$\Gamma_{\Mav,\mu}$ & $\gamma_5\gamma_{\mu}^{T}$ &$\gamma_5\frac{1}{2M_R}\sigma_{\mu\nu}P_\nu$  \\ 
\hline \hline
\end{tabular}
\end{center}
\vspace{-0.5cm}
\end{table}
\vspace{0cm}
For the first radial excitation of the $\Meps$ mesons, the matrix form of the BSE  (with ${\cal C}^{\rm iu}$ $\to$ ${\cal F}^{\rm iu}$ in the kernel), see~\cite{Gutierrez-Guerrero:2021rsx}, is:
\begin{equation}
\label{bsefinalEf}
\left[
\begin{array}{c}
E_{\Meps}(P)\\
\rule{0ex}{3.0ex} 
F_{\Meps}(P)
\end{array}
\right]
= \frac{4 \rmh}{3\pi}
\left[
\begin{array}{cc}
{\cal K}_{EE}^{\Meps} & {\cal K}_{EF}^{\Meps} \\
\rule{0ex}{3.0ex} 
{\cal K}_{FE}^{\Meps}& {\cal K}_{FF}^{\Meps}
\end{array}\right]
\left[\begin{array}{c}
E_{\Meps}(P)\\
\rule{0ex}{3.0ex} 
F_{\Meps}(P)
\end{array}
\right],  
\end{equation}
with
\begin{subequations}
\label{pionKernel}
\begin{eqnarray}
\nonumber
\nn {\cal K}_{EE}^{\Meps} &=&
\int_0^1d\alpha \bigg\{
{\cal F}^{\rm iu}(\omega^{(1)})  \\
\nn&&+ \bigg[ M_{\fu} M_{\fd}-\alpha (1-\alpha) P^2 - \omega^{(1)}\bigg]
\, \overline{\cal F}^{\rm iu}_1(\omega^{(1)})\bigg\},\\
\nn {\cal K}_{EF}^{\Meps} &=& \frac{P^2}{2 M_R} \int_0^1d\alpha\, \bigg[(1-\alpha)M_{\fu}+\alpha M_{\fd}\bigg]\overline{\cal F}^{\rm iu}_1(\omega^{(1)}),\\
\nn{\cal K}_{FE}^{\Meps} &=& \frac{2 M_R^2}{P^2} {\cal K}_{EF}^K ,\\
\nn {\cal K}_{FF}^{\Meps} &=& - \frac{1}{2} \int_0^1d\alpha\, \bigg[ M_{\fu} M_{\fd}+(1-\alpha) M_{\fu}^2+\alpha M_{\fd}^2\bigg] \\
\nn &\times &\overline{\cal F}^{\rm iu}_1(\omega^{(1)})\,,
\end{eqnarray}
\end{subequations}
where $\alpha$ is a Feynman parameter and we define the function $\omega^{(1)}\equiv\omega(M_{\fu}^2,M_{\fd}^2,\alpha,P^2)$ as
\begin{eqnarray}
\label{eq:omega}
\nn&& \hspace{-0.4cm} \omega^{(1)} =M_{\fu}^2 (1-\alpha) + \alpha M_{\fd}^2 + \alpha(1-\alpha) P^2\,.
\label{eq:C1}
\end{eqnarray} 
In analogy, we straightforwardly obtain the eigenvalue equations for $\Ms$, $\Mv$ and $\Mav$ excited states
\bea
\label{eig}
&&\nn 1-{\cal K}_{\Ms}(-m_{\Ms}^2)=0\\
&&\nn 1-{\cal K}_{\Mv}(-m_{\Mv}^2)=0\\
&& 1-{\cal K}_{\Mav}(-m_{\Mav}^2)=0\,,
\eea
where
\begin{eqnarray}\nn
\label{KastKernel} 
\nonumber  && \hspace{-2mm} 
{\cal K}_{\Ms}(P^2) = 
- \frac{4\rmh}{3\pi}
\int_0^1d\alpha\,\bigg[-{\cal L}_G
\overline{\cal F}_1^{\rm iu}(\omega^{(1)})\\
\nn && + \bigg({\cal F}^{\rm iu}(\omega^{(1)})
-{\cal F}_1^{\rm iu}(\omega^{(1)})\bigg) \bigg]\\
\nonumber && \hspace{-2mm} 
{\cal K}_{\Me}(P^2)= \frac{2\rmh}{3\pi} \int_0^1d\alpha\,
{\cal L}_{\Me}(P^2)
\overline{\cal F}_1^{\rm iu}(\omega^{(1)}),\\ 
\nn && \hspace{-4mm} {\cal K}_{\Mav}(P^2) =
\frac{2\rmh}{3\pi } \hspace{-2mm} \int_0^1 \hspace{-1mm} d\alpha \bigg[
{\cal F}_1^{\rm iu}(\omega^{(1)})
+ {\cal L}_{G}(P^2) \overline{\cal F}_1^{\rm iu}(\omega^{(1)})\bigg],\\
\end{eqnarray}
and we have defined
\bea  && \hspace{-4mm} \nn {\cal L}_{\Me}(P^2) \hspace{-1mm} = \hspace{-1mm} M_{\fu} M_{\fd} - (1-\alpha)M_{\fu}^2-\alpha M_{\fd}^2-2\alpha(1-\alpha)P^2,\\
&& \hspace{-4mm}\nn {\cal L}_{G}(P^2) \hspace{-1mm}=\hspace{-1mm}M_{\fd} M_{\fu}+\alpha (1-\alpha)P^2 \,.
\eea 
Eqs.~(\ref{bsefinalEf},\ref{eig}) have a physical solution when $P^2=-m_{H}^2$. Then the eigenvector 
corresponds to the BSA of the meson excitation.
It was observed in~\cite{Bermudez:2017bpx,Bashir:2011dp,Chang:2010hb,Chang:2011ei,Chang:2010jq} that the DCSB generates a large dressed-quark anomalous chromomagnetic moment and consequently the spin-orbit splitting between ground-state mesons and their parity partners is enhanced in accordance with observation.
In our study, we assume and implement this effect also in the parity partners between radial excitations of the mesons~\cite{Roberts:2011cf,Roberts:2011wy}.
This is the mechanism responsible for a magnified splitting between parity partners; namely, there are essentially nonperturbative DCSB corrections to the rainbow-ladder kernels, which largely cancel in the pseudoscalar and vector channels but add constructively in the scalar and axial-vector channels.
Following~\cite{Roberts:2011cf}, we introduce a spin-orbit repulsion into the scalar- and axial-vector radial excitations  through the artifice of a phenomenological coupling $g_{SO} \leq 1$, inserted as a single, common factor multiplying the kernels defined in Eqs.~(\ref{bsefinalEf}),~(\ref{eig}).
$g_{SO}$ mimics the dressed-quark chromomagnetic moment 
in full QCD.
The  first numerical value of $g_{SO} = 0.24$ was used in ~\cite{Roberts:2011cf} and later slightly modified in Refs.~\cite{Yin:2021uom,Lu:2017cln,Gutierrez-Guerrero:2019uwa}. For mesons with $J^P=0^+\;,1^+$ we use 
\bea
g_{SO}^{0^+}=0.32\;,\;\;\;g_{SO}^{1^+}=0.25\;. 
\eea
The value $g_{SO}^{1^+}=0.25$ in the axial vector channel guarantees the effect of spin-orbit repulsion and reproduces the desirable experimental value for the $a_1-\rho$ mass-splitting \cite{Yin:2021uom,Lu:2017cln,Gutierrez-Guerrero:2019uwa}.
On the other hand, $g_{SO}^{0^+}=0.32$ is chosen
to produce a mass difference of approximately $0.3$ GeV
between the quark-core of the $0{^+}(\tu\bar\td)$ (which we call  $\sigma(\tu\bar\td)$) and that of the $\rho$-meson (as obtained with beyond-RL kernels). The choice of 
$g_{SO}= 1$  indicates no repulsion and no additional interaction beyond that generated by the rainbow-ladder kernel.

\begin{figure}[H]
\vspace{-0.5cm}
\caption{Plotted $d_F$ obtained with Eq.~(\ref{fitdf}) and constants in Table~\ref{ec.fits} for first radial excitation of $\Meps$, $\Ms$, $\Mv$ and $\Mav$  mesons.}
       \centerline{\includegraphics[scale=0.3,angle=-90]{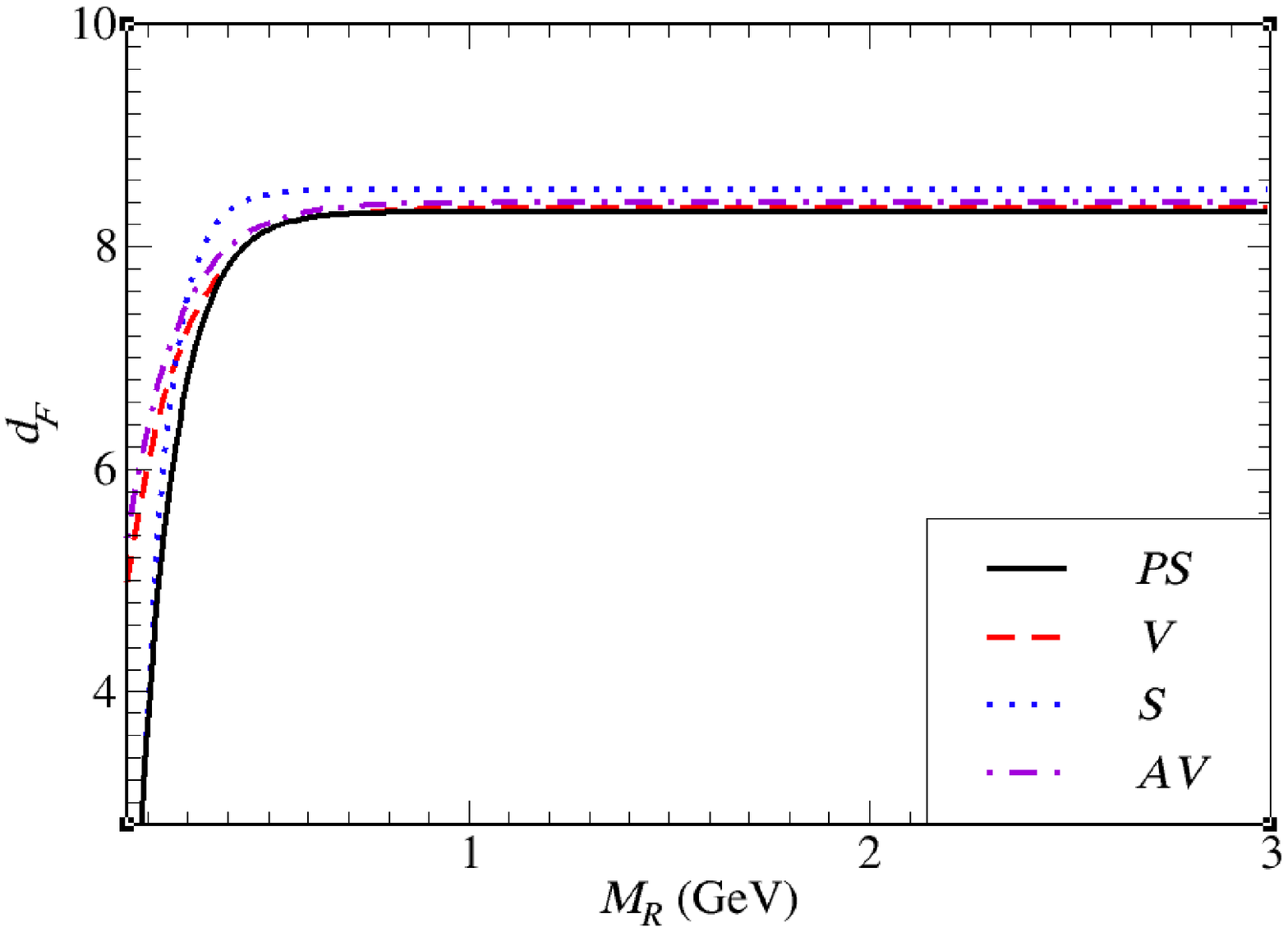}}
       \vspace{0cm}
       \label{Fig1}
\end{figure}

Now we only need to discuss how to choose the location of the zero in the excitation. For this purpose, it is necessary to fix the parameter $d_F$; this value was set to $d_F=2M^2$ in~\cite{Roberts:2011cf,Roberts:2011wy} 
for the calculation of radial excitations of light mesons in $\Meps$ and $\Mv$ channels. However, for heavy and heavy-light mesons, this value requires to be modified.
Our analysis for radial excitations of different types of mesons ($\Meps$, $\Ms$, $\Mv$, $\Mav$) composed for heavy or light quarks, shows that the best choice of $d_F$ can be adjusted fairly well with just one functional form:
\bea
\label{fitdf}
d_{\textit{F}}=d_1 - d_2 e^{-d_3 M_R} \,,
\eea
The constants $d_1, d_2, d_3$ for the different channels are displayed in Table~\ref{ec.fits} and $M_R$ is the reduced mass of the two quarks.
\begin{table}[htbp]
\caption{\label{ec.fits}
Selected parameters $d_1, d_2, d_3$ 
in Eq.~(\ref{fitdf}) for the first radial excitations of mesons.} 
\begin{center}
\begin{tabular}{@{\extracolsep{0.6 cm}}c|ccc}
\hline
\hline
Meson & $d_1$ & $d_2$ & $d_3$ \\
\hline
\rule{0ex}{2.5ex}
$\Meps$ & 8.32     & 41.67      & 11.08\\
\rule{0ex}{2.5ex}
$\Ms$ & 8.52     & 109.47     &15.82  \\
\rule{0ex}{2.5ex}
$\Mv$ & 8.35     & 10.37      & 7.50      \\
\rule{0ex}{2.5ex}
$\Mav$ & 8.44     & 10.00      &8.00     \\
\hline
\hline
\end{tabular}
\end{center}
\end{table}
In Fig.~\ref{Fig1}, we have plotted the curves that correspond to each of the fits for $d_F$. 
To obtain the parameters shown in Table~\ref{ec.fits}, we use existing experimental values for the masses of the excited states. Given the limited availability of experimental results, we supplement this information with the one obtained through using sophisticated theoretical models (see references in Table~\ref{ec.fits}) whenever possible. Guided by these experimental and theoretical results, we determined 
the best fit function $d_F(M_R)$~(\ref{fitdf}), such that it minimizes the associated average error for the predicted meson masses. We readily observe that the four curves lie in close proximity to each other, which suggests that it is possible to obtain a unified treatment of the first radial excitations of all $\Meps$, $\Ms$, $\Mv$ and $\Mav$  mesons.

Table~\ref{fbo} details the results obtained with the parameters of Tables~\ref{parameters},~\ref{table-M} and~\ref{ec.fits}. It also shows a comparison with the ground states, the experimental results and predictions obtained using other approaches.
\begin{table*}[ht] 
    \centering
    \caption{\label{fbo}Pseudoscalar and Scalar meson masses calculated from the BSE defined by Eqs. (\ref{bsefinalEf}) and (\ref{KastKernel}), using the parameter values in Tables.  \ref{parameters}, \ref{table-M} and \ref{ec.fits}. The experimental values are
taken from Ref \cite{Zyla:2020zbs}. Columns six and twelve show the percentage difference between the values predicted by our model and the experimental results. $n$ is the principal quantum number. $n = 0$ corresponds to the ground state and $n = 1$ to the first radial excitation.}
    \renewcommand{\arraystretch}{1.6} %
        \begin{tabular}{@{\extracolsep{0.4 cm}}cc|cccc|cc|cccc}
        \hline
        \multicolumn{6}{c|}{Pseudoscalar} & \multicolumn{6}{c}{Scalar}\\
        \hline
        \hline
        & $n$ & Exp. & Others & CI & Diff. \%  &  & $n$ & Exp. &  Others & CI & Diff. \%   \\  
         \hline
         \multirow{2}{*}{$\pi(\tu\bar{\td})$} & $0$ & $0.139$ & - & $0.14$ & $0.72$ &
         \multirow{2}{*}{$\sigma(\tu\bar{\td})$} & $0$  & $1.2$ & - & $1.22$ & $1.66$  \\
             & $1$ & $1.3$ & - & $1.27$ & $2.3$ &     & $1$ & - & $1.358$ \cite{Qin:2011dd} & $1.34$ & -  \\
             \hline
         \multirow{2}{*}{$K(\ts\bar{\tu})$} & $0$ & $0.493$ & - & $0.49$ & $0.61$  &  
         \multirow{2}{*}{$K_0^{*} (\tu\bar{\ts})$} & $0$ & $1.430$ & - & $1.33$ & $6.99$   \\
             & $1$ & $1.46$ & - & $1.51$ & $3.43$ &      & $1$ & - & $1.53$ \cite{Chen:2012qr}  & $1.57$ & -   \\
             \hline
         \multirow{2}{*}{$h_{\ts} (\ts\bar{\ts})$} & $0$ & - & - & $0.69$ & -  &  
         \multirow{2}{*}{$f_0 (\ts\bar{\ts})$} & $0$  & - & - & $1.34$ & -   \\
             & $1$ & - & - & $1.72$ & -  &      & $1$ & - & - & $1.82$ & -  \\
         \hline
         \multirow{2}{*}{$D^0(\tc\bar{\tu})$} & $0$ & $1.86$ & $1.88$ \cite{Godfrey:2015dva} & $1.87$ & $0.54$  &
         \multirow{2}{*}{$D_0^{*}(\tc\bar{\tu})$} & $0$  & $2.30$ & $2.45$ \cite{Godfrey:2015dva}  & $2.32$ & $0.87$  \\
             & $1$ & $2.54$ \cite{BaBar:2010zpy} & $2.58$ \cite{Godfrey:2015dva} & $2.53$ & $0.39$ &       & $1$ & - & $2.924$ \cite{Godfrey:2015dva} & $2.63$ & -   \\
         \hline
         \multirow{2}{*}{$D^{+}_{\ts}(\tc\bar{\ts})$} & $0$ & $1.97$ & $1.98$ \cite{Godfrey:2015dva} & $1.96$ & $0.51$ &  
         \multirow{2}{*}{$D^{*}_{\ts 0}(\tc\bar{\ts})$} & $0$  & $2.317$ & $2.55$ \cite{Godfrey:2015dva} & $2.43$ & $4.88$ \\
             & $1$ & - & $2.67$ \cite{Godfrey:2015dva} & $2.78$ & - &    & $1$ & $3.044$ \cite{BaBar:2009rro} & $3.18$ \cite{Godfrey:2015dva} & $3.27$ & $7.42$   \\
         \hline
         \multirow{2}{*}{$B^+ (\tu\bar{\tb})$} & $0$ & $5.28$ & $5.28$ \cite{Lu:2016bbk} & $5.28$ & $0$ &  
         \multirow{2}{*}{$B_0^{*}(\tu\bar{\tb})$} & $0$  & - & $5.72$ \cite{Lu:2016bbk} & $5.50$ & - \\
             & $1$ & $5.86$ \cite{LHCb:2015aaf} & $5.91$ \cite{Lu:2016bbk} & $5.68$ & $3.07$ &     & $1$ & - & $6.185$ \cite{Lu:2016bbk} & $5.82$ & - \\
         \hline
         \multirow{2}{*}{$B^0_{\ts}(\ts\bar{\tb})$} & $0$ & $5.37$ & $5.36$ \cite{Lu:2016bbk} & $5.37$ & $0$ & 
         \multirow{2}{*}{$B_{\ts 0}(\ts\bar{\tb})$} & $0$  & - & $5.80$ \cite{Lu:2016bbk} & $5.59$ & -   \\
             & $1$ & - & $5.98$ \cite{Lu:2016bbk} & $5.94$ & -      & & $1$ & - & $6.241$ \cite{Lu:2016bbk} & $6.57$ & -  \\
         \hline
        \multirow{2}{*}{$\eta_{\tc}(\tc\bar{\tc})$} & $0$ & $2.98$ & $2.93$ \cite{Fischer:2014cfa} & $2.98$ & $0$  &  
         \multirow{2}{*}{$\chi_{\tc 0}(\tc\bar{\tc})$} & $0$  & $3.414$ & $3.32$ \cite{Fischer:2014cfa} & $3.35$ & $1.87$   \\
             & $1$ & $3.64$ & $3.68$ \cite{Fischer:2014cfa} & $3.66$ & $0.55$ &       & $1$ & - & $3.83$ \cite{Fischer:2014cfa} & $4.7$ & -  \\
         \hline
         \multirow{2}{*}{$B^{+}_{\tc}(\tc\bar{\tb})$} & $0$ & $6.27$ & $6.27$ \cite{Li:2019tbn} & $6.28$ & $0.16$ &
         \multirow{2}{*}{$B_{\tc 0}(\tc\bar{\tb})$} & $0$  & - & $6.76$ \cite{Li:2019tbn} & $6.45$ & - \\
             & $1$ & $6.87$ & $6.87$ \cite{Li:2019tbn} & $6.80$ & $1.02$ &      & $1$ & -  & $7.134$ \cite{Li:2019tbn} & $6.88$ & -  \\
         \hline
         \multirow{2}{*}{$\eta_{\tb}(\tb\bar{\tb})$} & $0$ & $9.40$ & $9.41$ \cite{Fischer:2014cfa} & $9.40$ & $0$  &  
         \multirow{2}{*}{$\chi_{\tb 0}(\tb\bar{\tb})$} & $0$  & $9.859$ & $9.815$ \cite{Fischer:2014cfa} & $9.50$ & $3.64$   \\
             & $1$ & $9.99$ & $9.99$ \cite{Fischer:2014cfa} & $9.68$ & $3.10$ &     & $1$ & $10.232$ & $10.254$ \cite{Fischer:2014cfa} & $10.234$ & $0.02$ \\
         \hline
         \hline
    \end{tabular}
\end{table*}
We adopt the spectroscopic notation $n\,^{2S+1}L_J$, with $n$ the principal quantum number, $S$ the spin, $L$ the orbital angular momentum and $J$ the total angular momentum. 
In this notation, $n=0$ corresponds to the ground state and $n=1$ to the first radial excitation. 
We thus have:
\begin{eqnarray}
\begin{tabular}{@{\extracolsep{0.4 cm}}ccc}
 \nn {\Meps} & $\rightarrow$ & $n$\;$^{1}S_0$    \\
 \nn   {\Ms} &  $\rightarrow$ & $n$\;$^{1}P_1$  \\
  \nn   {\Mv} &  $\rightarrow $ & $n$\;$^{3}S_1$  \\
   \nn  {\Mav} & $\rightarrow$  & ( $n$\;$^{3}P_0$\;, $n$\;$^{3}P_1$\;, $n$\;$^{3}P_2$\;) 
   \end{tabular}
\end{eqnarray}
For pseudoscalar radial excitation $(1\;^{1}S_0)$ the largest difference between the results obtained with our model and the experimental data is $3.43\%$ which corresponds to the meson $K(\ts\bar{\tu})$ and the smallest difference is $0.39\%$ for the $D^0(\tc\bar{\tu})$ meson, while for vector mesons, the minimum and maximum differences are found in  $D^{*0}(\tc\bar{\tu})$ and $\phi (\ts\bar{\ts})$ mesons: $1.53\%$ and $6.55\%$ respectively. 
 We now check the experimental relation obtained in the CMS experiment at $\sqrt{s}=13\textmd{ TeV}$~\cite{CMS:2019uhm}
\bea M_{B^+_{\tc}(1S)} - M^{\text{rec}}_{B^{*}_{\tc}(1S)} =29\; \text{MeV} \,,
\eea
where
\bea M^{\text{rec}}_{B^{*}_{\tc}(1S)} = M_{B^{*}_{\tc}(1S)} - (M_{B^{*}_{\tc}(0S)}-M_{B^+_{\tc}(0S)}) \,.
\eea
In our CI model, this mass splitting is $20$ MeV which, being a difference of similar masses, is fairly consistent
with the measurement CMS measurement.

\begin{table*}[htbp]
    \centering
    \renewcommand{\arraystretch}{1.6} 
    \caption{\label{fbo2} Vector and Axial Vector meson masses calculated from the BSE defined by Eq. (\ref{KastKernel}), using the parameter values in Tables.  \ref{parameters}, \ref{table-M} and \ref{ec.fits}. The experimental values are
taken from Ref \cite{Zyla:2020zbs}. Columns six and twelve show the percentage difference between the values predicted by our model and the experimental results. $n$ is the principal quantum number and $n = 0$ corresponds to the ground state and $n = 1$ is the first radial excitation.} 
        \begin{tabular}{@{\extracolsep{0.2 cm}}cc|cccc|cc|cccc}
        \hline
        \multicolumn{6}{c|}{Vector} & \multicolumn{6}{c}{Axial Vector}\\
        \hline
        \hline
        & $n$ & Exp. & Others & CI & Diff. \%  &  & $n$ & Exp. &  Others & CI & Diff. \%   \\  
         \hline
         \multirow{2}{*}{$\rho(\tu\bar{\td})$} & $0$ & $0.78$ & - & $0.93$ & $19.23$ &   
         \multirow{2}{*}{$a_1(\tu\bar{\td})$} & $0$  & $1.260$ & - & $1.37$ & $8.73$   \\
             & $1$ & 1.47 & - & $1.47$ & 0  &      & $1$ & $1.65$ & - & $1.58$ & $4.01$   \\
         \hline
         \multirow{2}{*}{$K_1 (\tu\bar{\ts})$} & $0$ & $0.89$ & - & $1.03$ & $15.73$ &   
         \multirow{2}{*}{$K_1 (\tu\bar{\ts})$} & $0$  & $1.34$ & - & $1.48$ & $10.44$ \\
             & $1$ & $1.68$ & - & $1.63$ & $2.98$  &      & $1$ & - & $1.57$ \cite{Chen:2012qr} & $1.72$ & -  \\
         \hline
         \multirow{2}{*}{$\phi (\ts\bar{\ts})$} & $0$ & $1.02$ & - & $1.12$ & $9.80$  &  
         \multirow{2}{*}{$f_1 (\ts\bar{\ts})$} & $0$  & $1.43$ & - & $1.58$ & $10.49$   \\
             & $1$ & $1.68$ & - & $1.79$ & $6.55$  &      & $1$ & - & $1.67$ \cite{Chen:2012qr} & $1.88$ & -  \\
         \hline
         \multirow{2}{*}{$D^{*0} (\tc\bar{\tu})$} & $0$ & $2.01$ & $2.04$ \cite{Godfrey:2015dva} & $2.06$ & $2.49$ &  
         \multirow{2}{*}{$D_1 (\tc\bar{\tu})$} & $0$  & $2.420$ & $2.5$ \cite{Godfrey:2015dva} & $2.41$ & $0.41$  \\
             & $1$ & $2.61$ \cite{BaBar:2010zpy} & $2.64$ \cite{Godfrey:2015dva} & $2.57$ & $1.53$ &      & $1$ & - & $2.931$ \cite{Godfrey:2015dva} & $2.63$ & -   \\
         \hline
          \multirow{2}{*}{$D^*_{\ts} (\tc\bar{\ts})$} & $0$ & $2.11$ & $2.12$ \cite{Godfrey:2015dva} & $2.14$ & $1.42$  &  
         \multirow{2}{*}{$D_{\ts 1}(\tc\bar{\ts})$} & $0$  & $2.460$ & $2.6$ \cite{Godfrey:2015dva} & $2.51$ & $2.03$ \\
             & $1$ & $2.70$ & $2.73$ \cite{Godfrey:2015dva} & $2.78$ & $2.96$   &      & $1$ & - & $3.005$ \cite{Godfrey:2015dva} & $2.90$ & -   \\
         \hline
         \multirow{2}{*}{$B^{+*} (\tu\bar{\tb})$} & $0$ & $5.33$ & $5.33$ \cite{Lu:2016bbk} & $5.33$ & $0$ &
         \multirow{2}{*}{$B_1 (\tu\bar{\tb})$} & $0$  & $5.721$ & $5.77$ \cite{Lu:2016bbk} & $5.55$ & $2.99$  \\
             & $1$ & $5.97$ \cite{LHCb:2015aaf} & $5.94$ \cite{Lu:2016bbk} & $5.68$ & $4.86$ &      & $1$ & - & $6.145$ \cite{Lu:2016bbk} & $5.74$ & -  \\
         \hline
         \multirow{2}{*}{$B^{0*}_{\ts} (\ts\bar{\tb})$} & $0$ & $5.42$ & $5.41$ \cite{Lu:2016bbk} & $5.41$ & $0.18$  &  
         \multirow{2}{*}{$B_{\ts 1}(\ts\bar{\tb})$} & $0$  & $5.830$ & $5.85$ \cite{Lu:2016bbk} & $5.64$ & $3.26$  \\
             & $1$ & - & $6.0$ \cite{Lu:2016bbk} & $5.91$ & -  &      & $1$ & - & $6.2013$ \cite{Lu:2016bbk} & $6.05$ & -   \\
         \hline
         \multirow{2}{*}{$\Jpsi(\tc\bar{\tc})$} & $0$ & $3.10$ & $3.11$ \cite{Fischer:2014cfa} & $3.15$ & $1.61$  &  
         \multirow{2}{*}{$\chi_{\tc 1}(\tc\bar{\tc})$} & $0$  & $3.510$ & $3.49$ \cite{Fischer:2014cfa} & $3.40$ & $3.13$   \\
             & $1$ & $3.686$ & $3.7$ \cite{Fischer:2014cfa} & $3.92$ & $6.35$  &      & $1$ & - & $3.67$ \cite{Fischer:2014cfa} & $4.19$ & -  \\
         \hline
         \multirow{2}{*}{$B^{*}_{\tc}(\tc\bar{\tb})$} & $0$ & $6.27$ \cite{LHCb:2019bem} & $6.33$ \cite{Li:2019tbn} & $6.32$ & $0.80$ &
         \multirow{2}{*}{$B_{\tc \tb}(\tc\bar{\tb})$} & $0$  & - & $6.71$ \cite{Li:2019tbn} & $6.48$ & - \\
             & $1$ & $6.84$ \cite{LHCb:2019bem} & $6.89$ \cite{Li:2019tbn} & $6.85$ & $0.15$ &      & $1$ & - & $7.107$ \cite{Li:2019tbn} & $7.05$ & - \\
         \hline
         \multirow{2}{*}{$\Upsilon (\tb\bar{\tb})$} & $0$ & $9.46$ & $9.49$ \cite{Fischer:2014cfa} & $9.42$ & $0.42$ & 
         \multirow{2}{*}{$\chi_{\tb 1}(\tb\bar{\tb})$} & $0$  & $9.892$ & $9.842$ \cite{Fischer:2014cfa} & $9.52$ & $3.76$   \\
             & $1$ & $10.023$ & $10.9$ \cite{Fischer:2014cfa} & $9.71$ & $3.12$  &      & $1$ & $10.255$ & $10.120$ \cite{Fischer:2014cfa} & $9.53$ & $7.07$  \\
          \hline
        \hline
    \end{tabular}
\end{table*}


The masses of the scalar and axial vector mesons can be found in the Table~\ref{fbo2}.
Although there are few experimental results in these two channels, it is immediate to see that the maximum difference between our prediction and the observed result is $7.42\%$, which corresponds to the scalar meson $D^*_{\ts 0} (\tc\bar{\ts})$.
The low percentages differences lead us to conclude that our predictions are in fairly good agreement with the experimental measurements.

\begin{figure*}[htpb]
\caption{Comparison between the masses of the pseudoscalar ground state mesons and their first radial excitations.}
       \label{Fig2}
\vspace{-3cm}
    \centerline{\includegraphics[scale=0.61,angle=-90]{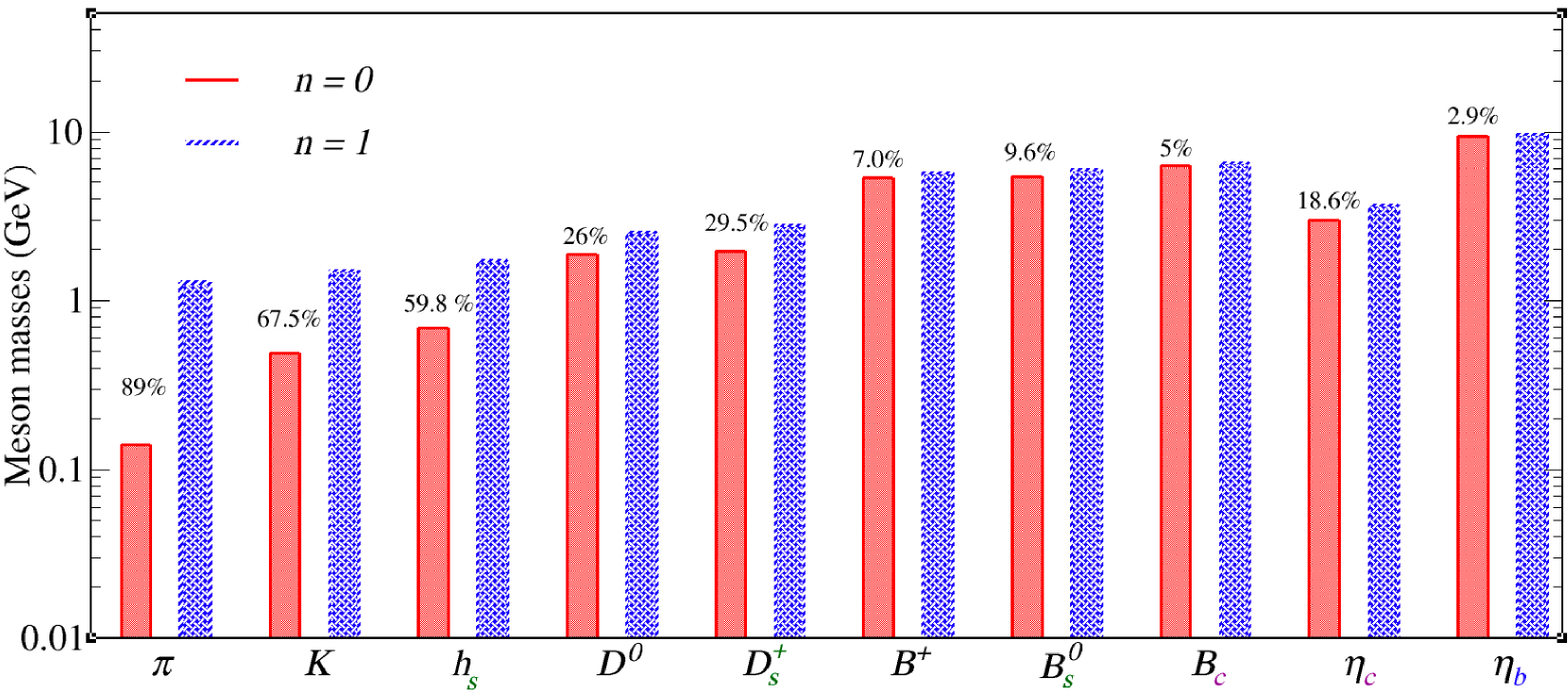}}
\vspace{-2cm}
\end{figure*}

\begin{figure*}[htpb]
\caption{Comparison between the masses of the ground states scalar mesons and their first radial excitations.}
       \label{Fig4}
       \vspace{-3cm}
       \centerline{\includegraphics[scale=0.61,angle=-90]{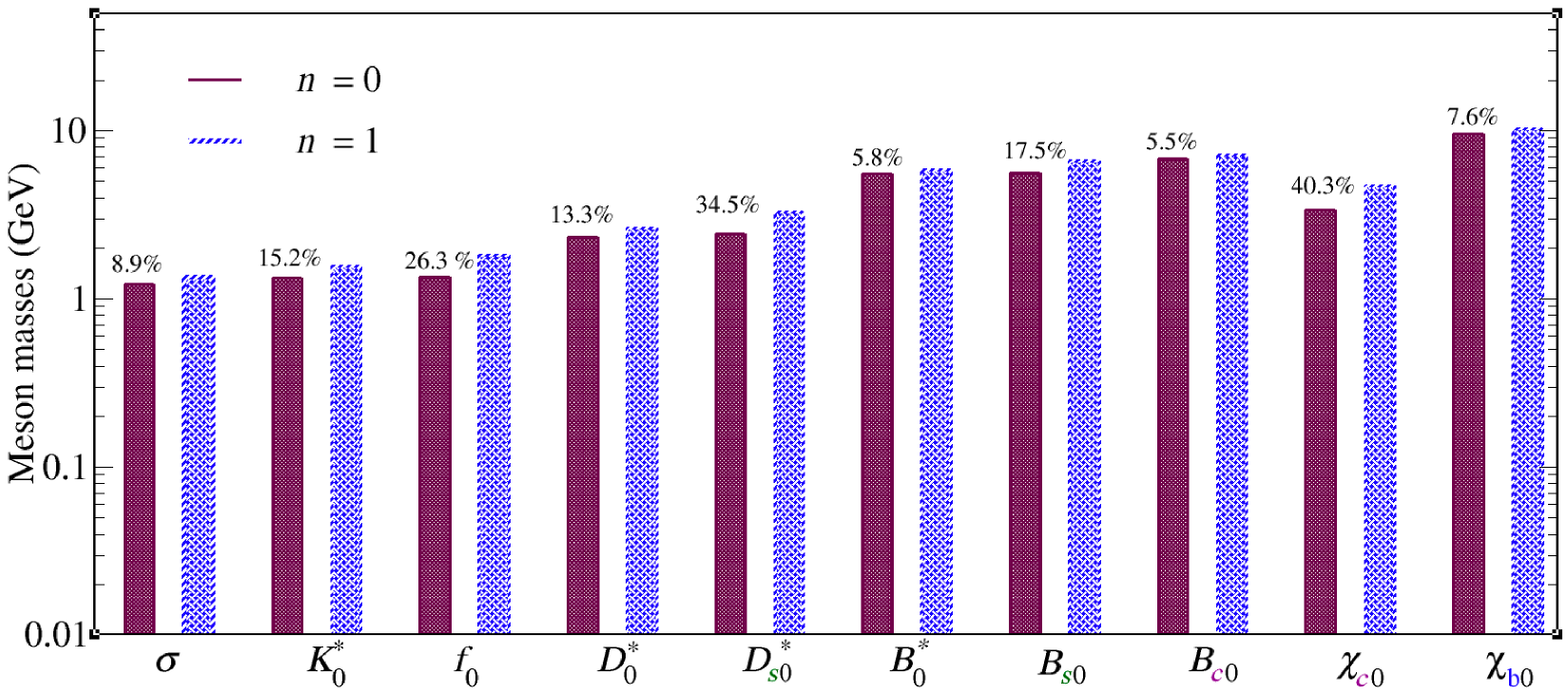}}
       \vspace{-3cm}
\end{figure*}

\begin{figure*}[htpb]
\caption{Comparison between the masses of the ground state vector mesons and their first radial excitations.}
       \label{Fig3}
\vspace{-3cm}
       \centerline{\includegraphics[scale=0.61,angle=-90]{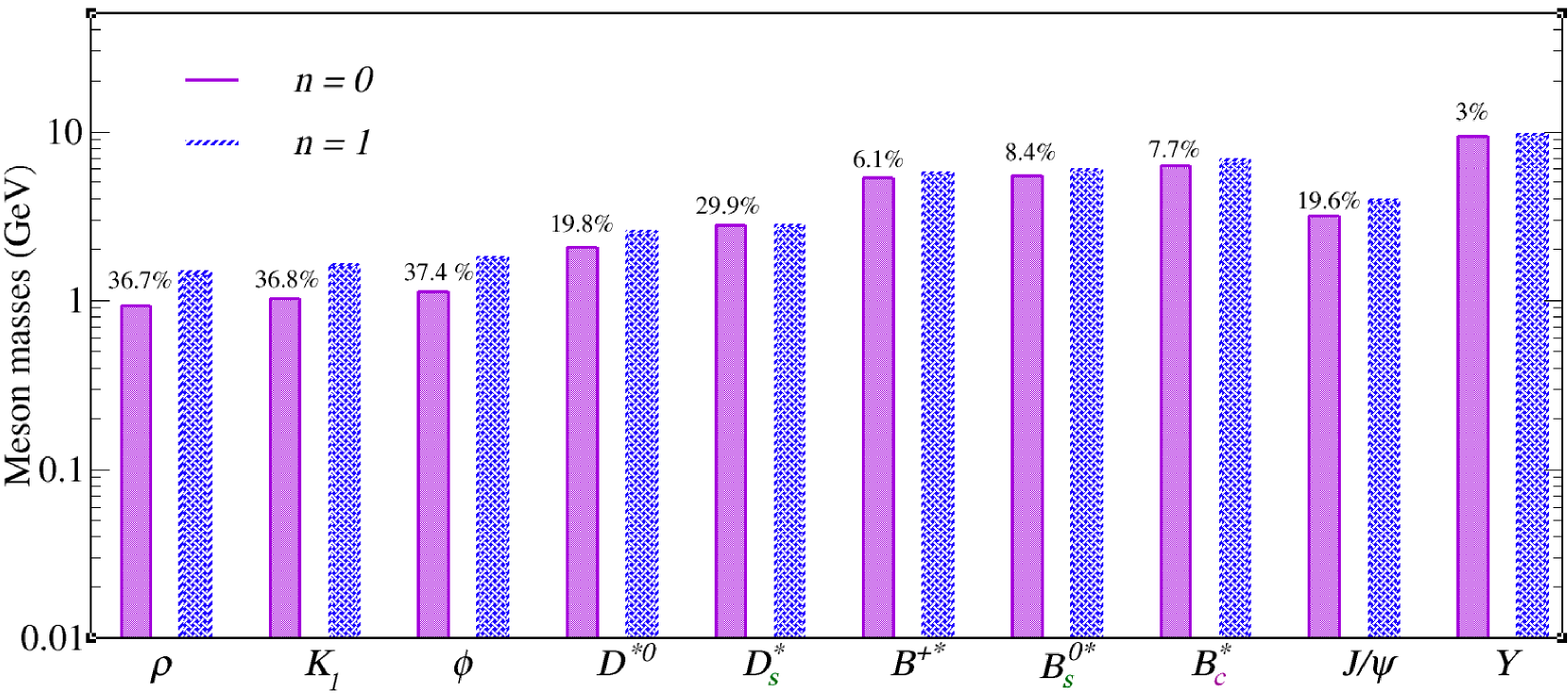}}
       \vspace{-2cm}
\end{figure*}

\begin{figure*}[htpb]
\caption{Comparison between the masses of the ground state axial vector mesons and their first radial excitations.}
\vspace{-3cm}
       \label{Fig5}
       \centerline{\includegraphics[scale=0.61,angle=-90]{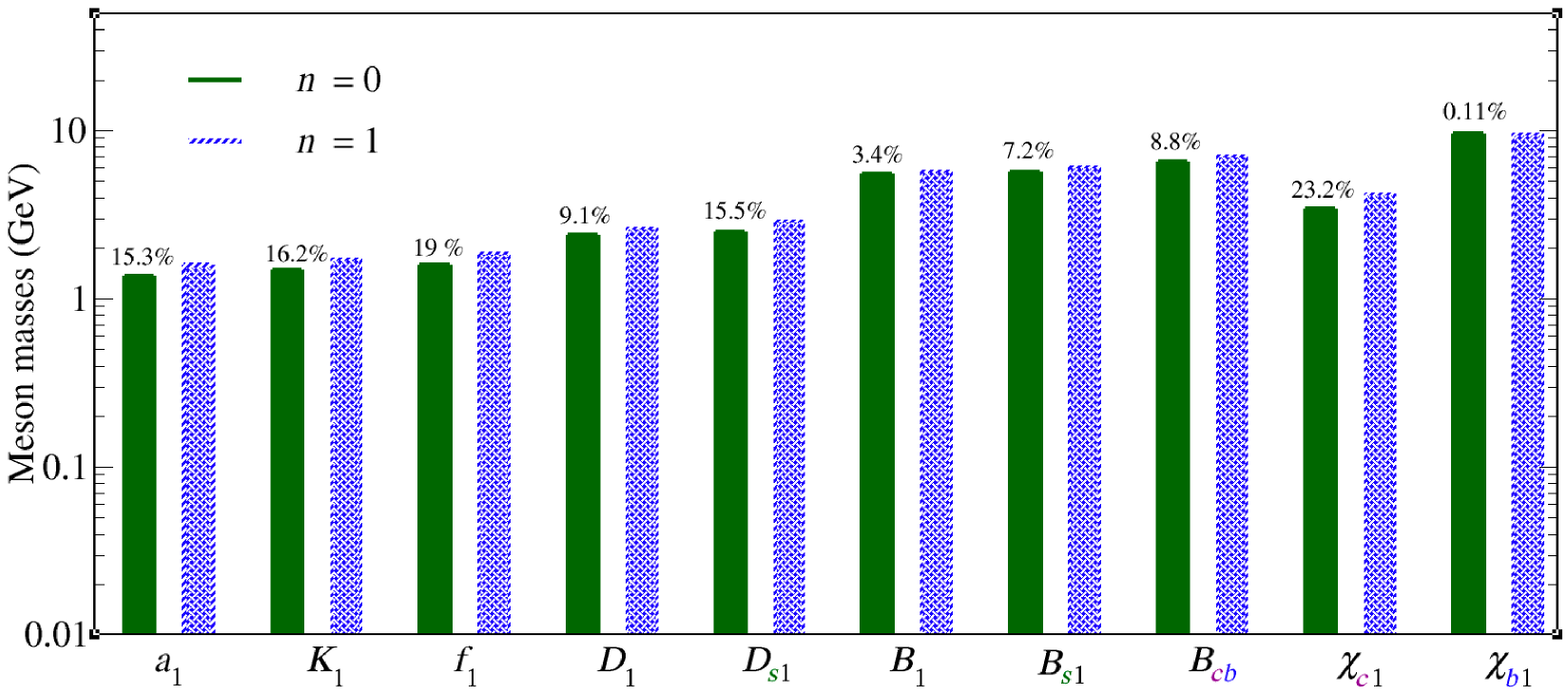}}
   \vspace{-3cm}
\end{figure*}

Figs.~\ref{Fig2},~\ref{Fig4},~\ref{Fig3} and~\ref{Fig5} provide a visual display of the difference in mass between the states $n = 0$ and $n = 1$ for $\Meps$, $\Ms$, $\Mv$ and $\Mav$  mesons.
Notably, the largest differences in mass is observed for light mesons. For states composed of one or two heavy quarks, this effect is less pronounced. In the case of the lightest meson, i.e. the pion, the difference is approximately $89\%$ and the heaviest meson $\chi_{\tb 1}(\tb\bar{\tb})$ has a difference of merely $0.11\%$ between the mass of its ground state and its first radial excitation.
Equal spacing rules for mesons \cite{Okubo:1961jc,GellMann:1962xb}  are obtained independently of the value of the quantum number $n$. Thus we expect them to be also valid for the radial excitations that we treat in our analysis. In that case, pseudoscalar and scalar mesons must satisfy the following mass relation for the radial excitations: \\ \\
$1\;^{1}S_0$: 
 \bea
 \nn
    m_{D_{\ts}^{+}(\tc\overline{\ts})} - m_{D^{0}(\tc\overline{\tu})} + m_{B^{+}(\tu\overline{\tb})} - m_{B_{\ts}^{0}(\ts\overline{\tb})} &=& 0 \,,\\ 
 \label{eq:1i} m_{\eta_{\tc(\tc\bar\tc)}} + m_{\eta_{\tb}(\tb \bar\tb)}- 2m_{B^+_{\tc}(\tc\bar\tb)} &=& 0 \,,
\eea  
$1\;^{1}P_1:$
\bea\label{eq:1ii}
    m_{D_{\ts0}^{*}(\tc\overline{\ts})} - m_{D^{*}_{0}(\tc\overline{\tu})} + m_{B^{*}_{0}(\tu\overline{\tb})} - m_{B_{\ts0}(\ts\overline{\tb})} &=& 0 \,. 
 \eea
 In our model, $\Meps$ mesons deviate  $1\%$ including the heaviest mesons ($\eta_{\tc}(\tc\bar\tc)$,  $\eta_{\tb}(\tb \bar\tb)$ and $B^+_{\tc}(\tc\bar\tb)$ ) and $\Ms$ mesons $11\%$ of Eqs.~(\ref{eq:1i},\ref{eq:1ii}). \\
Similarly, we can verify the equal spacing rules for the radial excitations of the ground state vector and axial vector mesons. For different combinations of light and heavy quarks, these relations read as: 

\noindent $1\;^{3}S_1:$
\bea 
\nn
    m_{D_{\ts}^{*}(\tc\overline{\ts})} - m_{D^{0*}(\tc\overline{\tu})} + m_{B^{+*}(\tu\overline{\tb})} - \nn m_{B_{\ts}^{0*}(\ts\overline{\tb})} &=& 0 \,,\\ \nn
 m_{\Jpsi(\tc\bar{\tc})} + m_{\varUpsilon(\tb\tb)} - 2m_{B^{*}_{c}(\tc\bar\tb)} &=& 0 \,,
\eea


$1\;^{3}P_0:$
\bea \label{eq:2}
    m_{D_{\ts1}(\tc\overline{\ts})} - m_{D_1(\tc\overline{\tu})} + m_{B_1(\tu\overline{\tb})} - m_{B_{\ts1}(\ts\overline{\tb})} &=& 0 \,.
    \label{GOM-2}
 \eea
We can check these relations by using the results presented in Table~\ref{fbo}. The deviation for vector heavy-light mesons is merely $2\%$ while for the heavy mesons comprised only with combinations of composing quarks $b$ and $c$: ($\Jpsi(\tc\bar{\tc})$,  $\varUpsilon(\tb\tb)$ and $B^{*+}_{c}(\tc\bar\tb)$ ), this difference still stays at a satisfactory $7\%$. Axial vector meson mass splitting difference in Eq.~(\ref{GOM-2}) is only at 3\%. With these results, we can verify that the mesons masses with mixed spin and parity and $n=1$ obey the following relations:
\bea \label{gmo-1}
&&\hspace{-0.7cm}m_{B_c^{*}(\tc\bar{\tb})}-m_{B_c^{+}(\tc\bar{\tb})} + m_{B_s^{0}(\ts\bar{\tb})}\approx m_{B_s^{0*}(\ts\bar{\tb})} \,,\\
\label{gmo-2}&& \hspace{-0.7cm}m_{B_{\ts}^{0*}(\ts\bar{\tb})}-m_{B^{+*}(\tu\bar{\tb})}+m_{B^{+}(\tu\bar{\tb})} = m_{B_{\ts}^0(\ts\bar{\tb})}\,,\\
\label{gmo-3}&&\hspace{-0.7cm}m_{B_{\ts}^{0*}(\ts\bar{\tb})}-m_{B^{+*}(\tu\bar{\tb})}+m_{D^{0}(\tc\bar{\tu})} = m_{D_{\ts}^{+}(\tc\bar{\ts})}\,, \\
\label{gmo-4}&&\hspace{-0.7cm}m_{\eta_{\tb}(\tb\bar{\tb})}-2m_{B_{\ts}^{0*}(\ts\bar{\tb})}+2m_{D_{\ts}^{*}(\tc\bar{\ts})}\approx m_{\eta_{\tc}(\tc\bar{\tc})} \,,\\
\label{gmo-5}&&\hspace{-0.7cm}m_{\eta_{\tb}(\tb\bar{\tb})}-2m_{B_{\ts}^0(\ts\bar{\tb})}+2m_{D_{\ts}^{+}(\tc\bar{\ts})}=m_{\eta_{\tc}(\tc\bar{\tc})}\,,\\
\label{gmo-6}&&\hspace{-0.7cm}m_{B_{\ts}^{0*}(\ts\bar{\tb})}-m_{D_{\ts}^{*}(\tc\bar{\ts})}+m_{D_{\ts}^{+}(\tc\bar{\ts})}= m_{B_{\ts}^0(\ts\bar{\tb})}\,,\\
\label{gmo-7}&&\hspace{-0.7cm}m_{\Upsilon(\tb\bar{\tb}) }-2m_{B_{\ts}^0(\ts\bar{\tb})}+2m_{D_{\ts}^{+}(\tc\bar{\ts})}= m_{\Jpsi(\tc\bar{\tc})}\,,\\
\label{gmo-8}&&\hspace{-0.7cm}m_{\Upsilon(\tb\bar{\tb}) }-m_{\eta_{\tb}(\tb\bar{\tb})}+m_{\eta_{\tc}(\tc\bar{\tc})}\approx m_{\Jpsi(\tc\bar{\tc})}\,,\\
\label{gmo-9}&&\hspace{-0.7cm}m_{\Upsilon(\tb\bar{\tb})} -2m_{B_{\ts}^{0*}(\ts\bar{\tb})}+2m_{D_{\ts}^{*}(\tc\bar{\ts})}\approx m_{\Jpsi(\tc\bar{\tc})}\,.
\eea

 The deviation from these mass relations is listed in Table~\ref{tab3:Resultados-reglas-espaciados-masas}, where we compare the result of each equation on the left-hand side with the result of the model on the right-hand side of the corresponding equation. We can readily conclude that our calculations agree very well with the mass relationships dictated by the Gell-Mann Okubo mass formulae~\cite{GellMann:1962xb,Okubo:1961jc} in all channels for the first radial excitations.

 This brings us to the conclusion of the detailed computation and presentation of our results for the first radial excitations of $\Meps$, $\Ms$, $\Mv$ and $\Mav$ mesons. We now take up diquarks in the next subsection.

 \begin{table*}[htbp]
\caption{The percentage difference (\%) of the equal spacing rules for the masses of mesons, Eqs.~(\ref{gmo-1}-\ref{gmo-9})  for CI.}
    \label{tab3:Resultados-reglas-espaciados-masas}
    \centering
        \begin{tabular}{@{\extracolsep{0.3 cm}} c  c  c  c  c  c  c  c  c  c}
            \hline\hline 
            &
            Eq. (\ref{gmo-1}) &
            \rule{0ex}{2.5ex}
            Eq. (\ref{gmo-2}) & 
            Eq. (\ref{gmo-3}) &
            \rule{0ex}{2.5ex}
            Eq. (\ref{gmo-4}) & 
            \rule{0ex}{2.5ex}
            Eq. (\ref{gmo-5}) & 
            \rule{0ex}{2.5ex}
            Eq. (\ref{gmo-6}) & 
            \rule{0ex}{2.5ex}
            Eq. (\ref{gmo-7}) & 
            \rule{0ex}{2.5ex}
            Eq. (\ref{gmo-8}) & 
            \rule{0ex}{2.5ex}
            Eq. (\ref{gmo-9})   \\
            CI & $1.35$ & $0.5$ & $0.72$ & $6.56$ & $8.20$ & $0.5$ & $13.5$ &
            $5.87$& $11.99$\\
            \hline\hline
        \end{tabular}
\end{table*}
\subsection{ Radial excitations of diquarks}
\label{diquark-radial}
 The idea of diquarks was introduced in \cite{Ida:1966ev}. Our current understanding of diquarks is no-point like quark-quark correlations with finite spatial extent. Lately, diquarks have been the focus of immense interest since they  provide a means for understanding baryonic properties. Diquarks can exist in a color sextet or a color anti-triplet combination. The interaction is attractive for diquarks in a color anti-triplet while it is repulsive in the color sextet channel [7].
Furthermore, it is the diquark in a color anti-triplet that can couple with a
quark to form a color-singlet baryon. 
We will only consider the diquark
correlations in a color anti-triplet configuration~\cite{Cahill:1987qr}. 
We use the notations $H=\Ds, \Dps, \Dav, \Dv$ which correspond to scalar, pseudoscalar, axial vector and vector diquarks, respectively. The BSA for diquarks has the same form as Eq.~(\ref{BSA-mesones}), but the coefficients change according to  Table~\ref{BSA-diquarks}.
\begin{table}[htbp]
 \caption{\label{ff-BSE} Here we list BSA for diquarks.}
\begin{center}
\label{BSA-diquarks}
\begin{tabular}{@{\extracolsep{0.5 cm}}lccc}
\hline \hline
BSA & $A_H$ & $B_H$ \\
\rule{0ex}{3.5ex}
$\Gamma_{DS}$  &$i\gamma_5$ & $\frac{1}{2M_R}\gamma_5 \gamma\cdot P$  \\
\rule{0ex}{3.5ex}
$\Gamma_{DPS}$  & $I_D$ & -- &  \\
\rule{0ex}{3.5ex}
$\Gamma_{DAV,\mu}$  & $\gamma_{\mu}^{T}$ & $\frac{1}{2M_R}\sigma_{\mu\nu}P_\nu$  \\
\rule{0ex}{3.5ex}
$\Gamma_{DV,\mu}$ & $\gamma_5\gamma_{\mu}^{T}$ & $\frac{1}{2M_R}\sigma_{\mu\nu}P_\nu$  \\
\hline \hline
\end{tabular}
\end{center}
\end{table}
The colour factor for mesons and diquarks is different owing to the fact that diquarks are color antitriplets, not singlets. It ensures quarks have attractive correlation within a diquark in the $\bar 3$ representation just as in mesons though the strength of this attraction is less.
The eigenvalue equation in the case of the first radial excitation of scalar diquarks is:
\begin{equation}
\label{bsefinalE}
\left[
\begin{array}{c}
E_{\Ds}(P)\\
\rule{0ex}{3.0ex} 
F_{\Ds}(P)
\end{array}
\right]
= \frac{4 \rmh}{6\pi}
\left[
\begin{array}{cc}
{\cal K}_{EE}^{\Meps} & {\cal K}_{EF}^{\Meps} \\
\rule{0ex}{3.0ex} 
{\cal K}_{FE}^{\Meps}& {\cal K}_{FF}^{\Meps}
\end{array}\right]
\left[\begin{array}{c}
E_{\Ds}(P)\\
\rule{0ex}{3.0ex} 
F_{\Ds}(P)
\end{array}
\right].
\end{equation}
The equations that will give us the masses of the pseudoscalar, axial vector and vector diquarks excitations are
\bea
\nn 0 & = & 1 + \frac{1}{2}{\cal K}_{\Ms}(-m_{\Dps}^2) \,,\\
\nn 0 & = & 1 - \frac{1}{2}{\cal K}_{\Mv}(-m_{\Dav}^2) \,,\\
0 & = & 1 + \frac{1}{2}{\cal K}_{\Mav}(-m_{\Dv}^2) \,.
\label{dq}\eea
From Eqs.~(\ref{bsefinalE}) and~(\ref{dq}) it follows that one may obtain the mass and BSA for a diquark with spin-parity $J^P$ from the equation for a $J^{-P}$ meson in which the only change is halving the interaction strength. The flipping of the sign in parity occurs because fermions and antifermions have opposite parity.
In Table~\ref{fbod}, we present our results for the masses of radially excited ($n=1$) diquark correlations  with $J^P=0^+,1^+,0^-,1^+$,  
compared to their ground state ($n=0$) masses.
In computing our results, we have made use of the 
function $d_F$ given by the expression~(\ref{fitdf}). In Table~\ref{fpor} we show the percentage difference between the states with $n =0$ and $n=1$.
As in the case of mesons, the mass of the first radial excitation is always greater than that of its ground state and as we expected, the percentage difference between these states is more noticeable in diquarks made up of light quarks. In fact, the difference in the vector diquark with two $\tb$ quarks is practically zero.

\begin{table}[htbp] 
    \centering
    \caption{\label{fbod}Diquark masses obtained
        using the parameters described in Table, \ref{parameters}, \ref{table-M} and the value of $d_F$ given by the fit in the expression (\ref{fitdf}) in the isospin symmetry limit. For pseudoscalar and vector diquarks, we show two results, one with $g_{SO}^{0^+}$ and $g_{SO}^{1^+}$ and the other with $g_{SO}^{0^+}$ and $g_{SO}^{1^+}$  multiplied by $1.8$ labeled with *.}
    \renewcommand{\arraystretch}{1.6} %
    \begin{tabular}{@{\extracolsep{0.4 cm}}cc|cccc}
         \hline
        \hline
        & $n$ & $\Ds$ & $\Dps$ & $\Dav$ & $\Dv$  \\  
         \hline
       \multirow{2}{*}{$\tu\td$} & $0$ & $0.77$ & $1.30$ $(1.15^*)$ & $1.06$ & $1.44$ $(1.33^*)$ \\
          & $1$  & $1.28$ & $1.39$ $(1.31^*)$ & $1.48$ & $1.62$ $(1.56^*)$  \\
         \hline
         \multirow{2}{*}{$\tu\ts$} & $0$ & $0.92$ & $1.41$ $(1.27^*)$ & $1.16$ & $1.54$ $(1.43^*)$  \\
          & $1$  & $1.52$ & $1.59$ $(1.56^*)$ & $1.63$  & $1.74$ $(1.70^*)$  \\
          \hline
         \multirow{2}{*}{$\ts\ts$}  & $0$ & $1.06$ & $1.51$ $(1.40^*)$ & $1.25$  & $1.64$ $(1.54^*)$  \\
          & $1$ & $1.72$ & $1.82$ $(1.81^*)$ & $1.80$  & $1.89$ $(1.87^*)$  \\
         \hline
        \multirow{2}{*}{$\tc\tu$} & $0$ & $2.08$ & $2.37$ $(2.28^*)$ & $2.16$  & $2.45$ $(2.38^*)$   \\
          & $1$  & $2.53$ & $2.64$ $(2.63^*)$ & $2.57$  & $2.65$ $(2.63^*)$ \\
        \hline
         \multirow{2}{*}{$\tc\ts$} & $0$ & $2.17$ & $2.47$ $(2.40^*)$ & $2.25$  & $2.54$ $(2.48^*)$\\
          & $1$  & $2.78$ & $3.27$ $(3.27^*)$ & $2.78$  & $2.90$ $(2.90^*)$ \\
        \hline
         \multirow{2}{*}{$\tu\tb$} & $0$ & $5.37$ & $5.53$ $(5.47^*)$ & $5.39$  & $5.59$ $(5.53^*)$ \\
          & $1$  & $5.68$ & $5.82$ $(5.82^*)$ & $5.68$  & $5.75$ $(5.74^*)$  \\
         \hline
         \multirow{2}{*}{$\ts\tb$} & $0$ & $5.46$ & $5.62$ $(5.57^*)$ & $5.47$  & $5.67$ $(5.62^*)$  \\
          & $1$  & $5.94$ & $6.50$ $(6.50^*)$  & $5.91$ & $6.05$ $(6.05^*)$ \\
         \hline
         \multirow{2}{*}{$\tc\tc$} & $0$ & $3.17$ & $3.38$ $(3.33^*)$ & $3.22$  & $3.42$ $(3.38^*)$ \\
          & $1$  & $3.90$ & $4.30$ $(4.30^*)$ & $3.92$ & $4.19$ $(4.19^*)$ \\
         \hline
        \multirow{2}{*}{$\tc\tb$} & $0$ & $6.35$ & $6.47$ $(6.44^*)$ & $6.35$  & $6.50$ $(6.47^*)$  \\
          & $1$  & $6.80$ & $7.07$ $(7.07^*)$ & $6.85$  & $7.05$ $(7.05^*)$ \\
        \hline
         \multirow{2}{*}{$\tb\tb$} & $0$ & $9.43$ & $9.51$ $(9.50^*)$ & $9.44$   & $9.53$ $(9.51^*)$ \\
          & $1$  & $9.68$ & $9.84$ $(9.84^*)$ & $9.71$  & $9.54$ $(9.52^*)$ \\
        \hline
        \hline
    \end{tabular}
    
\end{table}

\begin{table*}[htbp] 
    \centering
    \caption{\label{fpor}Percentage difference between the ground states of the diquarks and their first radial excitations. The masses are taken from the Table~\ref{fbod}.}
    \renewcommand{\arraystretch}{1.6} %
        \begin{tabular}{@{\extracolsep{0.8 cm}}ccccccccccc}
        \hline
        \hline
   & $\tu\td$ &
     $\tu\ts$&
      $\ts\ts$ &
      $\tc\tu$ &
      $\tc\ts$&
     $\tu\tb$ &
     $\ts\tb$ &
      $\tc\tc$ &
      $\tc\tb$ &
      $\tb\tb$\\
      $\Ds$ & $39.84$ & $39.47$ & $38.37$ & $17.78$ & $21.94$ & $5.45$ & $8.08$ & $18.71$ & $6.61$ & $2.58$\\
      $\Dps$ & $6.47$ & $11.32$ & $17.03$ & $10.22$ & $24.46$ & $4.98$ & $13.53$ & $21.39$ & $8.48$ &$3.35$\\
      $\Dav$ & $28.37$  & $28.83$  & $30.55$ & $15.95$  &  $19.06$  &  $5.10$  & $7.44$  & $17.85$  & $7.29$  & $2.78$\\
      $\Dv$ & $11.11$ & $11.49$ & $13.22$ & $7.54$ & $12.41$ & $2.78$ & $6.28$ & $18.37$ & $7.80$ & $0.01$\\
    \hline
    \hline
    \end{tabular}
    
\end{table*}

Following Ref.~\cite{Lu:2017cln},  we have multiplied $g_{SO}$ by a factor 1.8 for pseudoscalar and vector diquarks . 
This modification generates less repulsion. Physically, this might be understood by acknowledging that valence-quarks within a diquark are more loosely correlated than the valence-quark and -antiquark pair in a bound-state meson. Consequently, spin-orbit repulsion in diquarks should be less pronounced than it is in the corresponding mesons.

\section{Conclusions}
\label{Conclusions}

Using a symmetry-preserving regularisation of a vector $\times$ vector CI, we compute masses of the first radially excited state of forty mesons in Tables~\ref{fbo} and~\ref{fbo2} and of forty diquarks in Table~\ref{fbod}, including states that contain one or two heavy quarks.
Our predictions for the masses of the mesons are in good agreement with the values obtained experimentally observed values whenever available.
The maximum difference between our results and the experimental observation is $7.42\%$.
In Figs.~\ref{Fig2},~\ref{Fig3},~\ref{Fig4} and~\ref{Fig5}, we show a comparison between the masses of the ground states and their corresponding excited states. 
Notice that with the diquark masses and amplitudes described herein, one can construct all Faddeev kernels associated with radial excitations of octet and decouplet baryons as well as their chiral partners. All this is for future.

Finally we wish to clarify that many of the results
reported here for the mesons and  diquarks in the ground state have been previously obtained by us in~\cite{Gutierrez-Guerrero:2021rsx}. We report these results here along with their radial excitations for the sake of direct and ready comparison.  
Together with the results obtained in the present work, we have calculated around 218 states using the formalism described here. As indicated before, our future work is intended to be focused on computing masses of the first radial excitations of baryons.

\begin{acknowledgements}
G.~Paredes-Torres, L.~X.~Gutiérrez-Guerrero and also A. S. Miramontes acknowledge the financial support provided to them by the National Council for Humanities, Science and Technology (CONAHCyT), Mexico, 
   through their programs: 1) {\em Beca de Posgrado en Mexico}, 2) {\em Investigadoras e Investigadores por México}, and 3) {\em Postdoctorados Nacionales por M\'exico}. This research was also supported by {\em Coordinaci\'on de la Investigaci\'on Cientifica (CIC)} of the University of Michoacan through Grant No. 4.10 and {\em Ayudas Beatriz Galindo,} Spain.
 \\
\end{acknowledgements}

\bibliography{ccc-a}
\end{document}